\documentclass[11pt]{article}
\usepackage[dvips]{color}
\usepackage{epsfig}
\usepackage{amsmath}
\usepackage{graphicx}

\begin{document}
\title{Particles Collision near Kerr-Sen Dilaton-Axion Black Hole}
\author{{Ujjal
Debnath\thanks{Email: ujjaldebnath@gmail.com ;
ujjal@iucaa.ernet.in}}\\
{\small {\em Department of Mathematics, Indian Institute of Engineering Science and Technology,}}\\
{\small {\em Shibpur, Howrah-711 103, India.}}} \maketitle

\begin{abstract}
Here we consider static, axially symmetric, rotating and charged
Kerr-Sen Dilaton-Axion black hole metric in generalized
Boyer-Lindquist coordinates as particle accelerators. We obtain
the geodesic motions of particle. We find the effective potential
and discuss the circular orbit of a particle. We investigate the
center of mass energy of two colliding neutral particles with
different rest masses falling from rest at infinity to near the
non-extremal horizons (event horizon and Cauchy horizon) and extremal
horizon of the Kerr-Sen  black hole. Analogous to the Compton process,
we discuss the collision of a particle and a massless photon. Finally,
we find the center of mass energy due to the collision of two photons
in the background of Kerr-Sen black hole.
\end{abstract}

\section{Introduction}
Due to astrophysical research, Ba$\tilde{n}$ados, Silk and West
(BSW) \cite{BSW} have proposed the mechanism for the collision of
two particles falling from rest at infinity into the horizons of
Kerr black hole and determined that the center of mass (CM) energy
in the equatorial plane ($\theta=\pi/2$) may be high in the
extremal black hole. Lake \cite{Lake1} found that the CM energy of
the particles diverges to infinity at the inner horizon of
non-extremal Kerr black hole. Further, Wei et al \cite{Wei}
investigated that the CM energy of the collision of the particles
around Kerr-Newmann black hole depends on the spin ($a$) and
charge ($Q$) parameters. A general review of the collision
mechanism is found in ref \cite{Harada}. After that Liu et al
\cite{Liu} demonstrated the CM energy of the collision of
particles near the horizons of Kerr-Taub-NUT black hole and then
Zakria et al \cite{Zak} investigated the CM energy of the
collision for two neutral particles with different rest masses in
the background of a Kerr-Newman-Taub-NUT black hole. Sharif et al
\cite{Sharif2} found the CM energy of the collision of the
particles near the horizons of the most general accelerating
Plebanski-Demianski black hole and shown that the CM energy
depends on the accelerating parameter and other parameters also.
Particle acceleration in charged dilaton black hole has been
investigated in ref \cite{Pradhan1}. The authors in \cite{Said}
studied the particle accelerations and collisions in the back
ground of a cylindrical black holes.\\

On the other hand, Abhas et al \cite{Abhas} studied the geodesic
study of regular Hayward black hole and its rotating version has
been discussed by Bambi et al \cite{Bambi}. The particle
acceleration near the regular Hayward and Bardeen black holes has
been analyzed by Pradhan \cite{Pradhan}. Also the CM energy of the
collision for charged particles near the Bardeen black hole has
been analyzed \cite{Hai}. After that it was demonstrated the CM
energy of the collision of two particles near horizon of the
rotating Hayward's black hole \cite{Amir} and rotating Bardeen
black hole \cite{G}. Recently we have also found the CM energy for
the rotating versions of modified Hayward and Bardeen black hole
\cite{PU}. Tursunov et al \cite{Tur} have studied the particle
accelerations and collisions of black string. Fernando \cite{Fer}
has studied the possibility of high CM energy of two particles
colliding near the infinite red-shift surface of a charged BH in
string theory. Till now, several authors
\cite{Jacob,Berti,Ban,Har,P2,P1,P4,P3,Igata,Hus,Shar,Zas,Wei1,Pour1,Grib,Ghosh,Sharif1}
have studied the CM energy of the collision of particles near the
horizons of black holes. Our main motivation is to investigate the
geodesic study of the static, axially symmetric, rotating and
charged Kerr-Sen Dilaton-Axion black hole \cite{Gy}. The CM energy
and the particles orbit are investigated for two colliding neutral
particles of different rest masses falling from infinity to near
the event horizon, Cauchy horizon and extremal horizon of Kerr-Sen
Dilaton-Axion black hole. Also the collision of a particle with
massless photon and photon-photon collision near Kerr-Sen black
hole are discussed and the concerned CM energy are calculated.
Finally we conclude the results for particle acceleration near
the Kerr-Sen Dilaton-Axion black hole.\\

\section{Equations of motion for a particle near Kerr-Sen Dilaton-Axion Black Hole}
In 1992, Sen \cite{Sen} has constructed a solution of the
classical equations of motion arising in the low-energy effective
field theory for heterotic string theory. This low energy
heterotic string black hole carries a finite amount of charge,
angular momentum and magnetic dipole moment. It could be produced
by twisting method and starting from a rotating Kerr black hole
having no charge. So, sometimes it is called twisted Kerr black
hole or Kerr-Sen black hole. In Kerr-Sen black hole, the
thermodynamic properties are discussed in \cite{Lar,Pra} and
analytical gravitational lensing are discussed in \cite{Gy,Gy1}.
Hidden symmetries, null geodesics and photon capture in the Sen
black hole have been studied in \cite{Hioki}. Thermodynamic and
astrophysical consequences of the Sen black hole have been studied
by several authors \cite{J1,J2,J3,J4,J5,J6,J7,J8,J9}. Here we
consider static, axially symmetric, rotating and charged Kerr-Sen
Dilaton-Axion black hole metric in generalized Boyer-Lindquist
coordinates given by \cite{Gy}
\begin{eqnarray*}
ds^{2}=-\left(1-\frac{2Mr}{\rho^{2}(r)}\right)dt^{2}+\frac{\rho^{2}(r)}{\Delta(r)}~dr^{2}
+\rho^{2}(r)d\theta^{2}-\frac{4Mrasin^{2}\theta}{\rho^{2}(r)}~dtd\phi
\end{eqnarray*}
\begin{equation}
+\left(r(r+r_{\alpha})
+a^{2}+\frac{2Mra^{2}sin^{2}\theta}{\rho^{2}(r)}
\right)sin^{2}\theta d\phi^{2}
\end{equation}
where
\begin{equation}
\Delta(r)=r(r+r_{\alpha})-2Mr+a^{2}~,
\end{equation}
\begin{equation}
\rho^{2}(r)=r(r+r_{\alpha})+a^{2}cos^{2}\theta
\end{equation}
Here $M$ is the mass of the black hole, $a$ is the angular
momentum per unit mass of the black hole ($=J/M$) and
$r_{\alpha}=\frac{Q^{2}}{M}$, where $Q$ is the charge of the black
hole. From the Kerr-Sen black hole metric, we observe that there
is no naked singularity. For $a=0$, the Kerr-Sen black hole
reduces to the Gibbons-Maeda-Garfinkle-Horowitz-Strominger (GMGHS)
black hole and for $r_{\alpha}=0$, we get Kerr black hole. Further
if both $a=0$ and $r_{\alpha}=0$ then it reduces to Schwarzschild
black hole. The horizon found by setting $\Delta(r)=0$, i.e.,
\begin{equation}
r_{\pm}=\left(M-\frac{Q^{2}}{2M}\right)\pm
\sqrt{\left(M-\frac{Q^{2}}{2M}\right)^{2}-a^{2}}
\end{equation}
Here $r_{+}$ is the outer/event horizon and $r_{-}$ is the
inner/Cauchy horizon. The horizons exist if $a\le
M-\frac{Q^{2}}{2M}$. The extremal limit of Kerr-Sen black hole
describes $r_{+}=r_{-}=a=M-\frac{Q^{2}}{2M}$. The ergosphere
occurs when coefficient of $dt^{2}$ is zero i.e., $g_{tt}=0$. So
the radius of the ergosphere is given by
\begin{equation}
r_{ergo\pm}=\left(M-\frac{Q^{2}}{2M}\right)\pm
\sqrt{\left(M-\frac{Q^{2}}{2M}\right)^{2}-a^{2}cos^{2}\theta}
\end{equation}
Since $cos^{2}\theta<1$, so we have $r_{ergo+}>r_{+}$ and
$r_{ergo-}<r_{-}$. Thus the ergospheres occurs inside the Cauchy
horizon or outside the event horizon. The angular momentum
$\Omega$ at the horizon is obtained by the killing vector
($\partial_{t}+\Omega\partial_{\phi}$) is null at the horizon
\cite{Sen} i.e.,
\begin{equation}
g_{tt}+2g_{t\phi}\Omega+g_{\phi\phi}\Omega^{2}=0
\end{equation}
which simplifies to
\begin{equation}
\Omega_{\pm}\equiv\Omega=\frac{a}{2Mr_{\pm}}=\frac{1}{2Ma}\left[\left(M-\frac{Q^{2}}{2M}\right)\mp
\sqrt{\left(M-\frac{Q^{2}}{2M}\right)^{2}-a^{2}} \right]
\end{equation}
Since $r_{+}>r_{-}$, so we have $\Omega_{+}<\Omega_{-}$. Thus the
angular momentum on the Cauchy horizon is always larger than the
event horizon. Now consider a particle exhibits the geodesic
motion in the Kerr-Sen Dilaton-Axion spacetime. Let
$U^{a}=(U^{t},U^{r},U^{\theta},U^{\phi})$ be the four velocity of
the particle. On the equatorial motion of the particle,
$\theta=\pi/2$, so $U^{\theta}=0$. Thus the metric admits only two
killing vectors $\partial_{t}$ and $\partial_{\phi}$. Using these
killing vectors, the energy $E$ and angular momentum $L$ can be
defined by \cite{Sharif2,Sharif1}
\begin{equation}
E=-g_{ab}(\partial_{t})^{a}U^{b}=-g_{tt}U^{t}-g_{t\phi}U^{\phi}
\end{equation}
and
\begin{equation}
L=g_{ab}(\partial_{\phi})^{a}U^{b}=g_{t\phi}U^{t}+g_{\phi\phi}U^{\phi}
\end{equation}
From these, we obtain the 4 velocity vector along $t$ and $\phi$
directions as in the following forms:
\begin{equation}\label{10}
U^{t}=\frac{1}{\Delta(r)}\left[A(r)E-\frac{2Ma}{r+r_{\alpha}}~L\right]
\end{equation}
and
\begin{equation}\label{11}
U^{\phi}=\frac{1}{\Delta(r)}\left[\frac{2Ma}{r+r_{\alpha}}E+\left(1-\frac{2M}{r+r_{\alpha}}
\right)L\right]
\end{equation}
where
\begin{equation}\label{12}
A(r)=r(r+r_{\alpha})+a^{2}+\frac{2Ma^{2}}{r+r_{\alpha}}~~\ge 0
\end{equation}
On the horizon $r_{\pm}$, we must have $A(r_{\pm})=4M^{2}r_{\pm}$.
Now use the normalizing condition:
\begin{equation}\label{13}
g_{ab}U^{a}U^{b}=\epsilon
\end{equation}
where $\epsilon=-1$ represents the timelike geodesic, $\epsilon=0$
represents the null geodesic and $\epsilon=+1$ represents
spacelike geodesic. Now we obtain the radial component of the
4-velocity as
\begin{equation}\label{14}
U^{r}=\pm \frac{1}{\sqrt{g_{rr}}}
\left[\epsilon-g_{tt}(U^{t})^{2}-2g_{t\phi}U^{t}U^{\phi}-g_{\phi\phi}(U^{\phi})^{2}
\right]^{\frac{1}{2}}
\end{equation}
which yields
\begin{equation}\label{15}
U^{r}=\pm\frac{1}{\sqrt{r(r+r_{\alpha})}}
\left[A(r)E^{2}-\frac{4Ma}{r+r_{\alpha}}~EL-\left(1-\frac{2M}{r+r_{\alpha}}
\right)L^{2}+\epsilon\Delta(r)\right]^{\frac{1}{2}}
\end{equation}
where $\pm$ indicate the radially outgoing and incoming geodesics
respectively. Since the timelike component of the 4-velocity
($U^{t}$) is greater than zero (causally connected), so we get
\begin{equation}
E\left(r(r+r_{\alpha})+a^{2}+\frac{2Ma^{2}}{r+r_{\alpha}}
\right)\ge \frac{2LMa}{r+r_{\alpha}}
\end{equation}
On the horizon ($\Delta(r)=0$), the above inequality implies
\begin{equation}
L\le \frac{2MrE}{a}=\frac{E}{\Omega}
\end{equation}
The upper bound of the specific angular momentum ($E/\Omega$) at
the horizon of the Kerr-Sen black hole is known as the critical
angular momentum and denoted by $L_{c}$ and so
$L_{c}=\frac{2MrE}{a}$ where $r$ satisfies $\Delta(r)=0$. Now the
effective potential can be determined by equation
\begin{equation}
\frac{1}{2}(U^{r})^{2}+V_{eff}(r)=\frac{1}{2}(E^{2}+\epsilon)
\end{equation}
where
\begin{equation}
V_{eff}(r)=\frac{1}{2}(E^{2}+\epsilon)-\frac{1}{2r(r+r_{\alpha})}
\left[A(r)E^{2}-\frac{4Ma}{r+r_{\alpha}}~EL-\left(1-\frac{2M}{r+r_{\alpha}}
\right)L^{2}+\epsilon\Delta(r)\right]
\end{equation}
We observe that $V_{eff}(r)\rightarrow 0$ as $r\rightarrow\infty$
but $V_{eff}(r)$ diverges near the center ($r=0$) of the Kerr-Sen
black hole. Assume that the radius of circular orbit is $r=r_{c}$.
So the conditions for the circular orbit are given by
\begin{equation}
V_{eff}(r_{c})=0,~~\left[\frac{dV_{eff}(r)}{dr}\right]_{r=r_{c}}=0.
\end{equation}
which implies two equations
\begin{equation}
A(r_{c})E^{2}-\frac{4Ma}{r_{c}+r_{\alpha}}~EL-\left(1-\frac{2M}{r_{c}+r_{\alpha}}
\right)L^{2}+\epsilon\Delta(r_{c})-r(r+r_{c})(E^{2}+\epsilon)=0
\end{equation}
and
\begin{eqnarray*}
4aM(3r_{c}+r_{\alpha})EL+((r_{c}+r_{\alpha})(2r_{c}+r_{\alpha})
-2M(3r_{c}+r_{\alpha}))L^{2}+r_{c}(r_{c}+r_{\alpha})^{2}(A'(r_{c})E^{2}+\epsilon\Delta'(r_{c}))
\end{eqnarray*}
\begin{equation}
-(r_{c}+r_{\alpha})(2r_{c}+r_{\alpha})(A(r_{c})E^{2}-\epsilon\Delta(r_{c}))=0
\end{equation}
From these equations, we may obtain
\begin{eqnarray*}
aM[a^{2}+(r_{c}+r_{\alpha})(3r_{c}+r_{\alpha})]EL+
[-4a^{2}M+(r_{c}+r_{\alpha})\{(2r_{c}+r_{\alpha})(r_{c}+r_{\alpha})-2M(3r_{c}+r_{\alpha})\}]L^{2}
\end{eqnarray*}
\begin{equation}
+2\epsilon
M[2a^{4}+r_{c}^{2}(r_{c}+r_{\alpha})^{2}+a^{2}(r_{c}^{2}-r_{\alpha}^{2})-4a^{2}M(2r_{c}+r_{\alpha})
]=0
\end{equation}
From the above equation we may obtain the radius ($r_{c}$) of the
circular orbit of a particle near the Kerr-Sen black hole.

\section{Center of Mass energy for two neutral particles near horizons}

We now determine the center of mass energy (CME) for two colliding
particles with rest masses $m_{1}$ and $m_{2}$ moving in the
equatorial plane. The four-momentum of the $i$-th particle is
given by \cite{Zak}
\begin{equation}
(p_{i})^{a}=m_{i}(U_{i})^{a}~,~~i=1,2,~~~a=t,r,\theta,\phi
\end{equation}
where $p_{i}$ is the four momentum of the $i$-th particle. The CME
for two colliding particles is given by
\begin{equation}
E_{cm}^{2}=-(p_{i})^{a}(p_{i})_{a}=-[m_{1}(U_{1})^{a}+m_{2}(U_{2})^{a}][m_{1}(U_{1})_{a}+m_{2}(U_{2})_{a}]
\end{equation}
We assume the geodesic motion of the particles trajectories are
timelike ($\epsilon=-1$), $g_{ab}(U_{i})^{a}(U_{i})^{b}=-1$, so
the above equation can be written as
\begin{eqnarray*}
E_{cm}^{2}=2m_{1}m_{2}\left[\frac{(m_{1}-m_{2})^{2}}{2m_{1}m_{2}}+\left(1-g_{ab}U_{1}^{a}U_{2}^{b}
\right)
\right]~~~~~~~~~~~~~~~~~~~~~~~~~~~~~~~~~~~~~~~~~~~~~~~~~~~~~~~~~~~~~~~~~
\end{eqnarray*}
\begin{equation}
=2m_{1}m_{2}\left[\frac{(m_{1}-m_{2})^{2}}{2m_{1}m_{2}}+\left(1-g_{tt}U_{1}^{t}U_{2}^{t}
-g_{t\phi}U_{1}^{t}U_{2}^{\phi}-g_{\phi
t}U_{1}^{\phi}U_{2}^{t}-g_{\phi\phi}U_{1}^{\phi}U_{2}^{\phi}-g_{rr}U_{1}^{r}U_{2}^{r}\right)
\right]
\end{equation}
which turns out to be
\begin{equation}\label{25}
\frac{E_{cm}}{\sqrt{2m_{1}m_{2}}}=\left[\frac{(m_{1}-m_{2})^{2}}{2m_{1}m_{2}}
+\frac{X(r)-\kappa Y(r)}{\Delta(r)} \right]^{\frac{1}{2}}
\end{equation}
where
\begin{equation}\label{26}
X(r)=\left[A(r)E_{1}E_{2}-\frac{2Ma}{r+r_{\alpha}}~(E_{1}L_{2}+E_{2}L_{1})-\left(1-\frac{2M}{r+r_{\alpha}}
\right)L_{1}L_{2}+\Delta(r)\right],
\end{equation}
\begin{equation}\label{27}
Y(r)=\sqrt{Y_{1}(r)Y_{2}(r)}
\end{equation}
with
\begin{equation}\label{28}
Y_{i}(r)=\left[A(r)E_{i}^{2}-\frac{4Ma}{r+r_{\alpha}}~E_{i}L_{i}-\left(1-\frac{2M}{r+r_{\alpha}}
\right)L_{i}^{2}-\Delta(r)\right],i=1,2
\end{equation}
Here $E_{i}$ and $L_{i}$ are respectively the specific energy and
the specific angular momentum of the $i$-th particle ($i=1,2$). We
take $\kappa=+1$  when the particles move in the same direction
and $\kappa=-1$ when the particles move in the opposite direction.
We observe that the CME is invariant if we interchange the
quantities $m_{1}\leftrightarrow m_{2}$, $E_{1}\leftrightarrow
E_{2}$ and $L_{1}\leftrightarrow L_{2}$. We shall consider the
collisions of particles near the non-extremal horizons (event
horizon and Cauchy horizon) and extremal Kerr-Sen black hole and
discuss the properties of CME.

\subsection{Collision at non-extremal horizons}

On the non-extremal horizons $r=r_{\pm}$, we know that
$\Delta(r)=0$, so from above expression of the CME, the numerator
of the last term of eq (\ref{25}) should be zero. Using
L'Hospital's rule, we can write
\begin{eqnarray*}
\left.\frac{E_{cm}}{\sqrt{2m_{1}m_{2}}}\right|_{r\rightarrow
r_{\pm}}= \left[\frac{(m_{1}-m_{2})^{2}}{2m_{1}m_{2}}
+\frac{X'(r)-\kappa Y'(r)}{\Delta'(r)}
\right]^{\frac{1}{2}}_{r\rightarrow r_{\pm}}
\end{eqnarray*}
\begin{equation}
= \left[\frac{(m_{1}-m_{2})^{2}}{2m_{1}m_{2}}
+\frac{X'(r_{\pm})-\kappa Y'(r_{\pm})}{2r_{\pm}+r_{\alpha}-2M }
\right]^{\frac{1}{2}}
\end{equation}
Here $X'(r_{\pm})$ and $Y'(r_{\pm})$ are given by
\begin{equation}
X'(r_{\pm})=(2r_{\pm}+r_{\alpha}-2M)-\frac{2ML_{1}L_{2}}{(r_{\pm}+r_{\alpha})^{2}}
+\frac{2aM(E_{1}L_{2}+E_{2}L_{1})}{(r_{\pm}+r_{\alpha})^{2}}+E_{1}E_{2}
\left(2r_{\pm}+r_{\alpha}-\frac{2a^{2}M}{(r_{\pm}+r_{\alpha})^{2}}
\right),
\end{equation}
\begin{equation}
Y'(r_{\pm})=\frac{1}{2}\sqrt{Y_{1}(r_{\pm})Y_{2}(r_{\pm}) }
\left(\frac{Y'_{1}(r_{\pm})}{Y_{1}(r_{\pm})}+\frac{Y'_{2}(r_{\pm})}{Y_{2}(r_{\pm})}
\right),
\end{equation}
\begin{eqnarray*}
Y_{i}(r_{\pm})=\left[\left(r_{\pm}(r_{\pm}+r_{\alpha})+a^{2}+\frac{2Ma^{2}}{r_{\pm}+r_{\alpha}}
\right)E_{i}^{2}-\frac{4Ma}{r_{\pm}+r_{\alpha}}~E_{i}L_{i}
\right.
\end{eqnarray*}
\begin{equation}
\left. -\left(1-\frac{2M}{r_{\pm}+r_{\alpha}}
\right)L_{i}^{2}-\left(r_{\pm}(r_{\pm}+r_{\alpha})+a^{2}-2Mr_{\pm}
\right)\right],i=1,2,
\end{equation}
\begin{equation}
Y'_{i}(r_{\pm})=(2M-2r_{\pm}-r_{\alpha})-\frac{2ML_{i}^{2}}{(r_{\pm}+r_{\alpha})^{2}}
+\frac{4aME_{i}L_{i}}{(r_{\pm}+r_{\alpha})^{2}}+E_{i}^{2}
\left(2r_{\pm}+r_{\alpha}-\frac{2a^{2}M}{(r_{\pm}+r_{\alpha})^{2}}
\right),i=1,2
\end{equation}

\subsection{Collision at extremal horizon}

The extremal horizon (when the event horizon coincides with Cauchy
horizon) in Kerr-Sen black hole is located at
$r=r_{+}=r_{-}=a=M-\frac{Q^{2}}{M}$. So near the extremal horizon,
\begin{equation}
\left.\frac{E_{cm}}{\sqrt{2m_{1}m_{2}}}\right|_{r\rightarrow a}=
\left[\frac{(m_{1}-m_{2})^{2}}{2m_{1}m_{2}} +\frac{X''(r)-\kappa
Y''(r)}{\Delta''(r)} \right]^{\frac{1}{2}}_{r\rightarrow
a}=\left[\frac{(m_{1}-m_{2})^{2}}{2m_{1}m_{2}}
+\frac{X''(a)-\kappa Y''(a)}{2} \right]^{\frac{1}{2}}
\end{equation}
Here $X''(a)$ and $Y''(a)$ are given by
\begin{equation}
X''(a)=2+\frac{4ML_{1}L_{2}}{(a+r_{\alpha})^{3}}
-\frac{4aM(E_{1}L_{2}+E_{2}L_{1})}{(a+r_{\alpha})^{3}}+E_{1}E_{2}
\left(2+\frac{4a^{2}M}{(a+r_{\alpha})^{3}} \right),
\end{equation}
\begin{equation}
Y''(a)=\frac{1}{4}\sqrt{Y_{1}(a)Y_{2}(a)}
\left[2\frac{Y''_{1}(a)}{Y_{1}(a)}+2\frac{Y''_{2}(a)}{Y_{2}(a)}
-\left(\frac{Y'_{1}(a)}{Y_{1}(a)}-\frac{Y'_{2}(a)}{Y_{2}(a)}
\right)^{2} \right],
\end{equation}
\begin{eqnarray*}
Y_{i}(a)=\left[\left(2a^{2}+ar_{\alpha}+\frac{2Ma^{2}}{a+r_{\alpha}}
\right)E_{i}^{2}-\frac{4Ma}{a+r_{\alpha}}~E_{i}L_{i} \right.
\end{eqnarray*}
\begin{equation}
\left. -\left(1-\frac{2M}{a+r_{\alpha}}
\right)L_{i}^{2}-\left(2a^{2}+ar_{\alpha}-2Ma
\right)\right],i=1,2,
\end{equation}
\begin{equation}
Y'_{i}(a)=(2M-2a-r_{\alpha})-\frac{2ML_{i}^{2}}{(a+r_{\alpha})^{2}}
+\frac{4aME_{i}L_{i}}{(a+r_{\alpha})^{2}}+E_{i}^{2}
\left(2a+r_{\alpha}-\frac{2a^{2}M}{(a+r_{\alpha})^{2}}
\right),i=1,2,
\end{equation}
\begin{equation}
Y''_{i}(a)=-2+\frac{4ML_{i}^{2}}{(a+r_{\alpha})^{3}}
-\frac{8aME_{i}L_{i}}{(a+r_{\alpha})^{3}}+E_{i}^{2}
\left(2+\frac{4a^{2}M}{(a+r_{\alpha})^{3}} \right),i=1,2
\end{equation}

\subsection{CME near the center of the black hole}

The CME near the center ($r=0$) of the Kerr-Sen black hole is
given by (using equations (\ref{25})-(\ref{28}))
\begin{equation}
\left.\frac{E_{cm}}{\sqrt{2m_{1}m_{2}}}\right|_{r\rightarrow 0}=
\left[\frac{(m_{1}-m_{2})^{2}}{2m_{1}m_{2}} +\frac{X(0)-\kappa
Y(0)}{a^{2}} \right]^{\frac{1}{2}}
\end{equation}
where
\begin{equation}
X(0)=\left[\left(a^{2}+\frac{2Ma}{r_{\alpha}}\right)E_{1}E_{2}
-\frac{2Ma}{r_{\alpha}}~(E_{1}L_{2}+E_{2}L_{1})-\left(1-\frac{2M}{r_{\alpha}}
\right)L_{1}L_{2}+a^{2}\right],
\end{equation}
\begin{equation}
Y(0)=\sqrt{Y_{1}(0)Y_{2}(0)}
\end{equation}
with
\begin{equation}
Y_{i}(0)=\left[\left(a^{2}+\frac{2Ma}{r_{\alpha}}\right)E_{i}^{2}
-\frac{4Ma}{r_{\alpha}}~E_{i}L_{i}-\left(1-\frac{2M}{r_{\alpha}}
\right)L_{i}^{2}-a^{2}\right],i=1,2,
\end{equation}
We observe that CME $E_{cm}$ for colliding of two particles is
{\it finite} near the center of the Kerr-Sen black hole. Now if
there is no angular momentum (i.e., $a=0$) or there is no charge
(i.e., $r_{\alpha}=0 \implies Q=0$) of the black hole, then we see
that the CME $E_{cm}$ {\it diverges} near the center of the black
hole. But it can be easily determine that
$E_{cm}\rightarrow\infty$ at $r\rightarrow\infty$.

\section{Collision of a particle with photon}

Due to the process of Hawking radiation of the black hole, the
massless photon coming from the radiation can naturally scatter an
infalling particle. This phenomena is analogous to the Compton
scattering process. It was originally introduced for a photon and
an electron. Now we take into consideration is an infalling
particle collides with an outgoing massless photon \cite{Hali}.
Let $U^{a}=(U^{t},U^{r},U^{\theta},U^{\phi})$ be the four velocity
of the particle. On the equatorial motion of the particle,
$\theta=\pi/2$, so $U^{\theta}=0$. The timelike geodesic of a
particle satisfies $g_{ab}U^{a}U^{b}=-1$. So the 4 velocity
vectors $U^{t}$ and $U^{\phi}$ along $t$ and $\phi$ directions are
same in equations (\ref{10}) and (\ref{11}). But the radial
component of 4 velocity will be (putting $\epsilon=-1$ in eq
(\ref{15}))
\begin{equation}
U^{r}=\pm\frac{1}{\sqrt{r(r+r_{\alpha})}}
\left[A(r)E^{2}-\frac{4Ma}{r+r_{\alpha}}~EL-\left(1-\frac{2M}{r+r_{\alpha}}
\right)L^{2}-\Delta(r)\right]^{\frac{1}{2}}
\end{equation}
where $E$ and $L$ are respectively the energy and the angular
momentum of the particle and $A(r)$ is given in eq (\ref{12}). Let
us assume $K^{a}=(K^{t},K^{r},K^{\theta},K^{\phi})$ be the four
velocity of the photon. On the equatorial motion of the photon,
$\theta=\pi/2$, so $K^{\theta}=0$. The null geodesic of the photon
satisfies \cite{Hali} $g_{ab}K^{a}K^{b}=0$. So the 4 velocity
vectors $K^{t}$, $K^{\phi}$ and $K^{r}$ along $t$, $\phi$ and $r$
directions are similar to the equations (\ref{10}), (\ref{11}) and
(\ref{15})) (with $\epsilon=0$) and given by
\begin{equation} \label{41}
K^{t}=\frac{1}{\Delta(r)}\left[A(r)E_{\gamma}-\frac{2Ma}{r+r_{\alpha}}~L_{\gamma}\right],
\end{equation}
\begin{equation}\label{42}
K^{\phi}=\frac{1}{\Delta(r)}\left[\frac{2Ma}{r+r_{\alpha}}E_{\gamma}+\left(1-\frac{2M}{r+r_{\alpha}}
\right)L_{\gamma}\right]
\end{equation}
and
\begin{equation}\label{43}
K^{r}=\pm\frac{1}{\sqrt{r(r+r_{\alpha})}}
\left[A(r)E_{\gamma}^{2}-\frac{4Ma}{r+r_{\alpha}}~E_{\gamma}L_{\gamma}-\left(1-\frac{2M}{r+r_{\alpha}}
\right)L_{\gamma}^{2}\right]^{\frac{1}{2}}
\end{equation}
where $E_{\gamma}$ and $L_{\gamma}$ are respectively the energy
and the angular momentum of the photon. We now determine the
center of mass energy (CME) for collision of a particle with rest
mass $m$ and a massless photon moving in the equatorial plane.
Assume that the four-momentum of the particle is $p^{a}=mU^{a}$
and the photon is $k^{a}=K^{a}$ ($a=t,r,\theta,\phi$). So the CME
of a Hawking photon and the infalling particle can be taken as
\cite{Hali}
\begin{equation}
E_{cm}^{2}=-(p^{a}+k^{a})^{2}=m^{2}-2mg_{ab}U^{a}K^{b}
\end{equation}
which yields
\begin{equation}
E_{cm}^{2}=m^{2}+2m~\frac{P(r)-\kappa Q(r)}{\Delta(r)}
\end{equation}
where
\begin{equation}
P(r)=\left[A(r)EE_{\gamma}-\frac{2Ma}{r+r_{\alpha}}~(EL_{\gamma}+LE_{\gamma})-\left(1-\frac{2M}{r+r_{\alpha}}
\right)LL_{\gamma}\right],
\end{equation}
\begin{equation}
Q(r)=\sqrt{Q_{1}(r)Q_{2}(r)}
\end{equation}
with
\begin{equation}
Q_{1}(r)=\left[A(r)E^{2}-\frac{4Ma}{r+r_{\alpha}}~EL-\left(1-\frac{2M}{r+r_{\alpha}}
\right)L^{2}-\Delta(r)\right],
\end{equation}
\begin{equation}
Q_{2}(r)=\left[A(r)E_{\gamma}^{2}-\frac{4Ma}{r+r_{\alpha}}~E_{\gamma}L_{\gamma}-\left(1-\frac{2M}{r+r_{\alpha}}
\right)L_{\gamma}^{2}\right]
\end{equation}

\section{Photon-photon collision}

Halilsoy and Ovgun \cite{Hali} stated that the colliding energetic
photons in quantum electrodynamics can transmute into particles,
so their analysis was entirely classical and hence they have
assumed only to the center of mass energy of the yield without
further specification. Motivated by their work, here we shall also
find the center of mass energy for the collision between two
photons in the background of Kerr-Sen black hole. Let us assume
$K_{i}^{a}=(K_{i}^{t},K_{i}^{r},K_{i}^{\theta},K_{i}^{\phi})$
($i=1,2$) be the four velocities of the two photons. On the
equatorial motion of the photon, $\theta=\pi/2$, so
$K_{i}^{\theta}=0$. The two photons satisfy the null geodesics in
opposite directions. The null geodesic of the two photons satisfy
\cite{Hali} $g_{ab}K_{i}^{a}K_{i}^{b}=0$. So the 4 velocity
vectors $K_{i}^{t}$, $K_{i}^{\phi}$ and $K_{i}^{r}$ along $t$,
$\phi$ and $r$ directions are similar to the equations (\ref{41}),
(\ref{42}) and (\ref{43}) and given by
\begin{equation}
K_{i}^{t}=\frac{1}{\Delta(r)}\left[A(r)E_{\gamma_{i}}-\frac{2Ma}{r+r_{\alpha}}~L_{\gamma_{i}}\right],
\end{equation}
\begin{equation}
K_{i}^{\phi}=\frac{1}{\Delta(r)}\left[\frac{2Ma}{r+r_{\alpha}}E_{\gamma_{i}}+\left(1-\frac{2M}{r+r_{\alpha}}
\right)L_{\gamma_{i}}\right]
\end{equation}
and
\begin{equation}
K_{i}^{r}=\pm\frac{1}{\sqrt{r(r+r_{\alpha})}}
\left[A(r)E_{\gamma_{i}}^{2}-\frac{4Ma}{r+r_{\alpha}}~E_{\gamma_{i}}L_{\gamma_{i}}-\left(1-\frac{2M}{r+r_{\alpha}}
\right)L_{\gamma_{i}}^{2}\right]^{\frac{1}{2}}
\end{equation}
where $E_{\gamma_{i}}$ and $L_{\gamma_{i}}$ ($i=1,2$) are
respectively the energy and the angular momentum of the $i$-th
photon. We now determine the center of mass energy (CME) for
collision of two photons moving in the equatorial plane. Assume
that the four-momentum of the photons are $k_{i}^{a}=K_{i}^{a}$
($a=t,r,\theta,\phi$; $i=1,2$). So the CME of the colliding
photons can be taken as \cite{Hali}
\begin{equation}
E_{cm}^{2}=-(k_{1}^{a}+k_{2}^{a})^{2}=-2g_{ab}K_{1}^{a}K_{2}^{b}
\end{equation}
which yields
\begin{equation}
E_{cm}^{2}=2~\frac{Z(r)-\kappa W(r)}{\Delta(r)}
\end{equation}
where
\begin{equation}
Z(r)=\left[A(r)E_{\gamma_{1}}E_{\gamma_{2}}-\frac{2Ma}{r+r_{\alpha}}~
(E_{\gamma_{1}}L_{\gamma_{2}}+E_{\gamma_{2}}L_{\gamma_{1}})
-\left(1-\frac{2M}{r+r_{\alpha}}
\right)L_{\gamma_{1}}L_{\gamma_{2}}\right],
\end{equation}
\begin{equation}
W(r)=\sqrt{W_{1}(r)W_{2}(r)}
\end{equation}
with
\begin{equation}
W_{i}(r)=\left[A(r)E_{\gamma_{i}}^{2}-\frac{4Ma}{r+r_{\alpha}}~E_{\gamma_{i}}L_{\gamma_{i}}
-\left(1-\frac{2M}{r+r_{\alpha}}
\right)L_{\gamma_{i}}^{2}\right],~i=1,2.
\end{equation}

\section{Discussions and Concluding Remarks}
In this work, we have considered static, axially symmetric,
rotating and charged Kerr-Sen Dilaton-Axion black hole metric in
generalized Boyer-Lindquist coordinates as particle accelerators.
We have obtained the geodesic motions of a particle. We found the
effective potential and also found the radius of the circular
orbit of a particle. We have investigated the center of mass
energy (CME) of two colliding neutral particles (which satisfied
timelike geodesic) with different rest masses $m_{1}$ and $m_{2}$
falling from rest at infinity to near the non-extremal horizons
(event horizon and Cauchy horizon) and extremal horizon of the
Kerr-Sen black hole. We observed that CME $E_{cm}$ for colliding
of two particles is {\it finite} near the center of the Kerr-Sen
black hole. Now if there is no angular momentum (i.e., $a=0$) or
there is no charge (i.e., $r_{\alpha}=0 \implies Q=0$) of the
black hole, then we have seen that the CME $E_{cm}$ {\it diverges}
near the center of the black hole. But we
found that $E_{cm}\rightarrow\infty$ at $r\rightarrow\infty$.\\

Analogous to the Compton process, we found the CME for the
collision of a particle (which satisfied timelike geodesic) with
mass $m$ and a massless photon (which satisfied null geodesic)
where photons can be coming from the Hawking radiation of the
black hole. Finally, we found the CME due to the collision of two
photons (which satisfied null geodesic) in the background of Kerr-Sen black hole.\\

{\bf Acknowledgement:} The author is thankful to
IUCAA, Pune, India for warm hospitality where the work was carried out.\\

\end{document}